\documentclass[twocolumn,superscriptaddress]{revtex4-2}
\usepackage[colorlinks=true,citecolor=blue,linkcolor=blue,urlcolor=blue,]{hyperref}
\usepackage{graphicx, nicefrac, textcomp}
\usepackage[normalem]{ulem}
\usepackage{changes}
\usepackage{chngcntr}
\usepackage{amssymb}
\usepackage{soul}
\usepackage{color}
\usepackage{float}
\usepackage{textcomp}
\usepackage{amstext}
\usepackage{graphicx}
\usepackage{gensymb}
\usepackage{graphics}\usepackage{subfigure}\usepackage{longtable}\usepackage{pstricks}\usepackage{dcolumn}\usepackage{bm}
\usepackage{siunitx}
\usepackage{wasysym}  
\begin{document}

 \title{Chemo~-~Structural Disorder in the kagom\'e spin $S$~=~1/2 systems ZnCu$_3$(OH)$_6$Cl$_2$ and  YCu$_3$(OH)$_{6}$Br$_{2}$[Br$_x$(OH)$_{1-x}$]}
\author{Reinhard K. Kremer}
\affiliation{Max-Planck-Institute for Solid State Research, Heisenbergstra{\ss}e 1, 70569 Stuttgart, Germany}
\author{Sebastian Bette}
\affiliation{Max-Planck-Institute for Solid State Research, Heisenbergstra{\ss}e 1, 70569 Stuttgart, Germany}
\author{J\"urgen Nuss}
\affiliation{Max-Planck-Institute for Solid State Research, Heisenbergstra{\ss}e 1, 70569 Stuttgart, Germany}
 \author{Pascal Puphal}
\email[]{puphal@fkf.mpg.de}
\affiliation{Max-Planck-Institute for Solid State Research, Heisenbergstra{\ss}e 1, 70569 Stuttgart, Germany}
\affiliation{2nd Physics Institute, University of
Stuttgart, 70569 Stuttgart, Germany}

\date{\today}

\begin{abstract}
By single crystal diffraction we characterize the chemo~-~structural disorder introduced by Zn~-~Cu site mixing in the kagom\'e spin $S$~-~1/2 systems herbertsmithite ZnCu$_3$(OH)$_6$Cl$_2$ and  YCu$_3$(OH)$_{6}$Br$_{2}$[Br$_x$(OH)$_{1-x}$]. For an untwinned single crystal of herbertsmithite of composition Zn$_{0.95(1)}$Cu$_{2.99(3)}$O$_{5.9(1)}$H$_{5.8(1)}$Cl$_2$ we find substitution by Cu of the Zn atoms in the layers separating the kagom\'e layers as well as substantial Zn substitution for Cu in the kagom\'e layers.
In  YCu$_3$(OH)$_{6}$Br$_{2}$[Br$_x$(OH)$_{1-x}$] site mixing disorder is present for intermediate $x$. Analogous to the Cl homologous system in crystals  with $x$~=~1/3 
disorder is absent and a low-temperature structural transition emerges driven by strong magneto-phonon coupling as a release of frustration. Apart from this structural anomaly we find the physical properties of these crystals unchanged compared to intermediate $x$ and closely resembling the Cl homologue where long-range magnetic order was observed.
\end{abstract}

\maketitle

\section{Introduction}
Low-dimensional materials with strong magnetic frustration, such as compounds containing decoupled antiferromagnetic kagom\'e layers, are prototypical candidates to search for an experimental realization of the quantum spin-liquid (QSL) phase \cite{Balents10,Chamorro2020}. This long-sought unusual quantum state exhibits no static magnetic order, despite sizable exchange interactions. Rather, macroscopic entanglement and fractional excitations are expected. 
Cu-based systems with perfect kagom\'e layers of Cu$^{2+}$ spin $S$~=~1/2 cations are considered prime QSL candidates and have been strongly investigated in the past.
One of the first systems for which clear experimental signatures of a QSL phase were observed is herbertsmithite with composition ZnCu$_3$(OH)$_6$Cl$_2$ \cite{mendels07,Han2012}. Lately, the  Zn-substituted barlowite, ZnCu$_3$(OH)$_6$BrF, was proposed as another putative QSL candidate \cite{Tustain20,Fu21}.

Herbertsmithite ZnCu$_3$(OH)$_6$Cl$_2$ represents the $x$~=~1 endmember of the substitution series of Zn$_x$Cu$_{4-x}$(OH)$_6$Cl$_2$ crystallizing in the space group $R\overline3m$ with ABC-type stacking  of the kagom\'e layers along the $c$ axis. The kagom\'e layers are well separated by triangular layers of nonmagnetic Zn-anions, centering non-distorted oxygen octahedra. The kagom\'e layers result by connecting strongly Jahn-Teller elongated CuO$_4$Cl$_2$ octahedra leaving the Cu$^{2+}$ cations with a square planar oxygen coordination and Cl$^-$ anions at the elongated apical sites of the  coordination octahedra.

Very early on chemical and structural disorder, for example, induced by  site mixing of magnetic and nonmagnetic ions, or deviations from the perfect kagom\'e arrangement  were a controversial topic  when discussing the magnetic properties of  herbertsmithite \cite{Imai2008,Rozenberg2008}.
By using neutron powder diffraction on a deuterated herbertsmithite sample assuming a Cu$^{2+}$ to Zn$^{2+}$ ratio of 3~:~1 and no vacant cation sites, de Vries \textit{et al.} reported a 91:9\% Cu~:~Zn occupancy of the kagom\'e sites corresponding to a 73\% occupancy of the interlayer Zn sites \cite{Vries2008,Vries2010}.
On the other hand, Freedman \textit{et al.} investigated a crystal with composition Zn$_{0.85}$Cu$_{3.15}$(O H)$_6$Cl$_2$ by anomalous x-ray scattering and detected Cu atoms at the Zn sites but concluded no Zn substitution of the Cu kagom\'e layer sites\cite{Freedman2010}. 
Lattice distortions due to a Jahn-Teller effect of the Cu$^{2+}$ ions substituting for the interlayer Zn$^{2+}$ cations were discussed by Han \textit{et al.} \cite{Han2016}.
By Cu$^{2+}$ electron spin resonance Zorko \textit{et al.} found evidence for magnetic defects whose response violates trigonal symmetry \cite{Zorko2017}. They even suggested a global structural distortion related to the formation of the spin-liquid ground state. Norman \textit{et al.} recently investigated possible global symmetry lowering in herbertsmithite and pointed out that small perturbations could stabilize an anisotropic spin-liquid or a valence bond solid \cite{Norman2020}.
As a consequence of site mixing often intricate sample dependence of  the magnetic properties of herbertsmithite samples is observed which  makes an unequivocal assessment of the ground state properties extremely challenging and, indeed, gapped \cite{Fu2015,Han2016} and nongapped ground states have been discussed \cite{Wulferding2010,Han2012,Fu2015,Khuntia2020}.

With A$^{n+}$Cu$_3$(OH)$_6$Cl$_n$ (n~=~2, 3) a new  series of related compounds has been found with A$^{3+}$Cu$_3$(OH)$_6$Cl$_3$~\cite{Puphal2018,Sun2016,Fu2021}, realizing perfect kagom\'e layers in a kapellasite-type structure (AA-stacking), the structural polymorph of herbertsmithite. For YCu$_3$(OH)$_{6+x}$Cl$_{3}$ long range magnetic order (LRO) below a transition temperature of 15~K was reported driven by a strong Dzyaloshinskii-Moriya (DM) interaction \cite{Barthelemy2019,Zorko2019,Zorko2019a,Prelovsek2021}.

However, for A~=~Y the compound is not stable against water, leading to Cl-OH substitution and phases with compositions of YCu$_3$(OH)$_{6+x}$Cl$_{3-x}$ are formed, until the endmember Y$_3$Cu$_9$(OH)$_{19}$Cl$_{8}$ for $x$~=~1/3 is obtained ~\cite{Puphal2017}. This substitutional variant shows a distortion of the  Cu lattice as Y atoms move out of the kagom\'e plane\cite{Puphal2017} creating two non-equivalent Cu sites and a unique magnetic model with three different nearest-neighbor exchange interactions is required\cite{Hering2022}. 
For Y$_3$Cu$_9$(OH)$_{19}$Cl$_{8}$ chemical disorder-free crystals can be obtained which undergo LRO below 2.2~K, as spin waves were seen in inelastic neutron scattering, and fast damped oscillations in $\mu$SR were detected \cite{Chatterjee2023}. Multimagnon modes were accessed by terahertz time-domain spectroscopy, further confirming LRO, and a 1/6 magnetization plateau was observed.\cite{Biesner2022} Finally, in  \textsuperscript{1}H-NMR a clear peak of the spin-lattice relaxation rate  $T^{-1}$ was observed at $T_{\rm N}$ \cite{Wang2023}.

Notably, a blend of the two Y phases with an intermediate $x$ can be synthesized where a partial occupancy of  the Y atoms is present\cite{Sun2016}. In Ref. \onlinecite{Sun2021} a mixture of the two phases was reported with a major fraction of the Y atoms out of the plane. Specific heat measurements revealed two distinct anomalies and an entropy release at both transitions \cite{Chatterjee2023}.
Recent low-temperature crystal structure investigations on various kagom\'e systems revealed structural instabilities which induce distortions as seen, for example, in the herbertsmithite parent compound clinoatacamite~\cite{Zheng2013,Malcherek2017}, as well as for barlowite, claringbullite~\cite{Henderson2019}, volborthite~\cite{Ishikawa2015} and vesigneite~\cite{Boldrin2016} indicating it to be a prevalent structural motif at low temperatures. Similarly, for  Y$_3$Cu$_9$(OH)$_{19}$Cl$_{8}$ crystals as evidenced by thermal expansion and specific heat measurements a 33~K structural anomaly has been seen, which is suppressed in powder samples \cite{Chatterjee2023}. These low temperature distortions are attributed to enhanced magnetoelastic coupling, if the energies of magnetic interactions and phonons coincide, as was shown, for Y$_3$Cu$_9$(OH)$_{19}$Cl$_{8}$ in detailed phonon investigations by ellipsometry studies \cite{Dolezal2024}.
In polycrystalline samples of Y$_3$Cu$_9$(OH)$_{19}$Cl$_{8}$  obtained by a solid state reaction route \cite{Barthelemy2019}, hence avoiding water in the preparation, disorder serves as chemical strain locally pinning the structure leading to the absence of structural transitions down to lowest temperatures \cite{Chatterjee2023}. External strain is, however, in comparison too weak to suppress LRO in this system \cite{Wang2023}. Interestingly, such disordered powder samples of Y$_3$Cu$_9$(OH)$_{19}$Cl$_{8}$ without the structural anomaly showed a completely different $\mu$SR spectrum and clear absence of LRO \cite{Barthelemy2019}.



In addition to the chlorides, the sister compound series YCu$_3$(OH)$_{6}$Br$_{2}$[Br$_x$(OH)$_{1-x}$] was also synthesized \cite{Chen2020}. Here, the crystal structure determination proposed a mixture of the two end members similar to the Cl system, with partial occupancy of both Y sites, within and out of the kagom\'e layers, similar to crystals described in Ref. \cite{Sun2021}.  Variation of $x$ has been reported \cite{zeng2022possible,Liu2022,Lu2022} even reaching the composition YCu$_3$(OH)$_{6.5}$Cl$_{2.5}$\cite{Chen2020}, which  exceeds the endmember $x=1/3$, i.e. YCu$_3$(OH)${19}$Br$_8$ \cite{Lu2022}.  Unfortunately, in most studies the crystal structures were not determined \cite{Hong2022,Zeng2024,Suetsugu2024}.
Recently, the Br system attracted particular attention due to signatures of a Dirac spin-liquid ground state \cite{Liu2022,zeng2022possible} based on a $T^2$ term in  specific heat and a Dirac cone shaped excitation spectrum found in inelastic neutron scattering experiments \cite{Zeng2024}. Thermal transport studies revealed thermally activated phonon-spin scattering indicating a gapped magnetic excitation spectrum \cite{Hong2022}. and 1/9 and 1/3 magnetization plateaus were found  \cite{Zheng2023,Suetsugu2024}. In addition, quantum oscillations in torque measurements are observed for the first time in an insulator \cite{Zheng2023}. 
However, $\mu$SR and NQR results revealed an inhomogeneous ground state with both fast and slow relaxing components \cite{Lee2024}.
Lately, by mixing Cl and Br the next level of complexity was reached~\cite{Li2024,Shivaram2024,Xu2024}.  

Here, by using single crystal XRD, we report on investigations of the chemical disorder in ZnCu$_3$(OH)$_6$Cl$_2$ and, in particular, we compare in detail single crystals of the systems YCu$_3$(OH)$_{6}$Cl$_{3}$, Y$_3$Cu$_9$(OH)$_{19}$Cl$_{8}$, and Y$_3$Cu$_9$(OH)$_{19}$Br$_{8}$. By single crystal XRD we find for the Br systems the same superstructure as for the Cl variants leading to a distorted kagom\'e lattice, with the absence of mixed O/Br and Y site disorder. The physical properties of these crystals are similar to those reported in literature, with the distinction that now in the specific heat an  anomaly at 15~K is observed which  is field independent and likely similar to the 33~K anomaly in the Cl variant.

\section{Experimental}
\subsection{Sample Preparation and Composition}
Single crystals of ZnCu$_3$(OH)$_6$Cl$_2$ were grown similar to the recipe described in detail in Ref. \onlinecite{Han2011} by placing a thick walled Pyrex glass ampoule ($\phi_{inner}$=15 mm, $\phi_{outer}$=21 mm, $l=150$ mm) filled with  0.2g of CuO, 2g ZnCl$_2$, and 4 ml H$_2$O in an oven with temperature gradient of 150 to 180$\degree$C. The exact composition of the crystal used in XRD was determined by ICP-MS and gas extraction and found to be Zn$_{0.95(1)}$Cu$_{2.99(3)}$O$_{5.9(1)}$H$_{5.8(1)}$Cl$_2$.

Crystals of YCu$_3$(OH)$_{6}$Cl$_{3}$ and Y$_3$Cu$_9$(OH)$_{19}$Cl$_{8}$ were grown by two different routes following the descriptions provided in Refs.~\onlinecite{Sun2016,Biesner2022,Chatterjee2023}: The first by a waterfree hydroflux variant mixing 0.6g LiOH, 1.5g BaCl$_2$, 2g Y(NO$_3$)$_3$-6H$_2$O,	1g CuCl$_2$-2H$_2$O, and 0.6g CuCl heated to 240$\degree$C for four days in a 23 ml teflon lined autoclave. The second by liquid phase transport of CuO in a YCl$_3$-H$_2$O solution at elevated temperatures of 240$\degree$C at the hot end of similarly thick walled glass ampoules as used in the preparation of hebertsmithite.

Crystals of Y$_3$Cu$_9$(OH)$_{19}$Br$_{8}$ were synthesized following Ref.~\cite{Chen2020}. 0.6 g Cu(NO$_3$)$_2$ $\cdot$ 3H$_2$O, 1.915 g Y(NO$_3$)$_3\cdot$ 6H$_2$O and 1.785 g KBr (molar ratio Cu : Y : Br = 1 : 2 : 6) were mixed and  1 mL deionized water was added. Without further homogenization the mixture was transferred to a 23 mL teflon lined autoclave and heated to 230$\degree$C for 3 days. 
By repeating the synthesis several times we noticed that KBr is slowly creeping through the teflon liner and hence the process differs from the Cl cas as crystallization is more a consequence of over-saturation and less of supercooling. Temperature variations confirmed this observation in the sense that higher temperatures led to a faster concentration change and different Br deficient phases were found. We tested the variation of the composition, and the temperature profile and noticed that, similar to the Cl system, it is important to  apply a temperature above 180$\degree$C as otherwise the parent compound Cu$_2$(OH)$_3$Br forms which undergoes a magnetic transition around 10~K~\cite{Zheng2009}. Optimized synthesis conditions were obtained with 0.6 g Cu(NO$_3$)$_2$ $\cdot$ 3H$_2$O 1.9g Y(NO$_3$)$_3\cdot$ 6H$_2$O, 2.2 g KBr in 2.2 mL deionized water heated to 230$\degree$C for 2 days and subsequently quenched to room temperature. 

\subsection{Single Crystal XRD}
XRD at room temperature was performed on single crystals of typical lateral sizes of 20 $\mu$m which had been broken off under high viscosity oil from larger items. The crystals were attached with vacuum grease on a loop made of Kapton foil (Micromounts, MiTeGen, Ithaca, NY). Diffraction data were collected with a SMART APEXI CCD X-ray diffractometer (Bruker AXS, Karlsruhe, Germany), using graphite-monochromated  Mo-K$_{\alpha}$ radiation ($\lambda$~=~0.71073~\AA).
The reflection intensities were integrated with the SAINT subprogram in the Bruker Suite software, a multi-scan absorption correction was applied using SADABS \cite{Sheldrick2015}. In case of Y$_3$Cu$_9$(OH)$_{19}$Cl$_{8}$ and Y$_3$Cu$_9$(OH)$_{19}$Br$_{8}$ the diffraction data show patterns of reverse/obverse twins (see Figures \ref{Y}b and c), typical for rhombohedral symmetry. The data were corrected for this type of twinning applying multi-scan absorption correction with TWINABS \cite{sheldrick2012twinned}. All crystal structures were solved by direct methods and refined by full-matrix least-square fitting with the SHELXTL software package \cite{sheldrick2008sadabs,Sheldrick2015}.

\subsection{Low temperature powder XRD}
Low temperature powder XRD have been carried out using a Bruker D8 Venture with Cu K$_\alpha^1$ radiation equipped with a closed-cycle He refrigerator. High statistic runs have been taken measuring in the range of 10 to 140$\degree$ for 13 h in the range of 50-12~K.

\subsection{Magnetization and Specific Heat}
Magnetization measurements were performed on single crystals at temperatures $0.4\le T \le 350$~K using a superconducting quantum interference device (MPMS-XL, Quantum Design) equipped with the He3 option. 
Specific-heat measurements were carried out, down to $T$ = 400~mK in a {Physical Properties Measurement System} (PPMS, Quantum Design) similarly equipped with the He3 option. 

\section{Results and Discussion}
\subsection{Herbertsmithite}
Herbertsmithite is one of the best investigated  QSL system with signatures of spinons seen in inelastic neutron scattering \cite{Han2012} and the clear absence of frozen spins in $\mu$SR \cite{mendels07}. As discussed above Zn:Cu site mixing can lead to structural and magnetic perturbations of the kagom\'e planes.

By single-crystal XRD we investigated an untwinned single crystal grown by the established recipe which according to chemical analysis had a composition of Zn$_{0.95(1)}$Cu$_{2.99(3)}$O$_{5.9(1)}$H$_{5.8(1)}$Cl$_2$.

\begin{figure}[h]
\centering
\includegraphics[width=1\columnwidth]{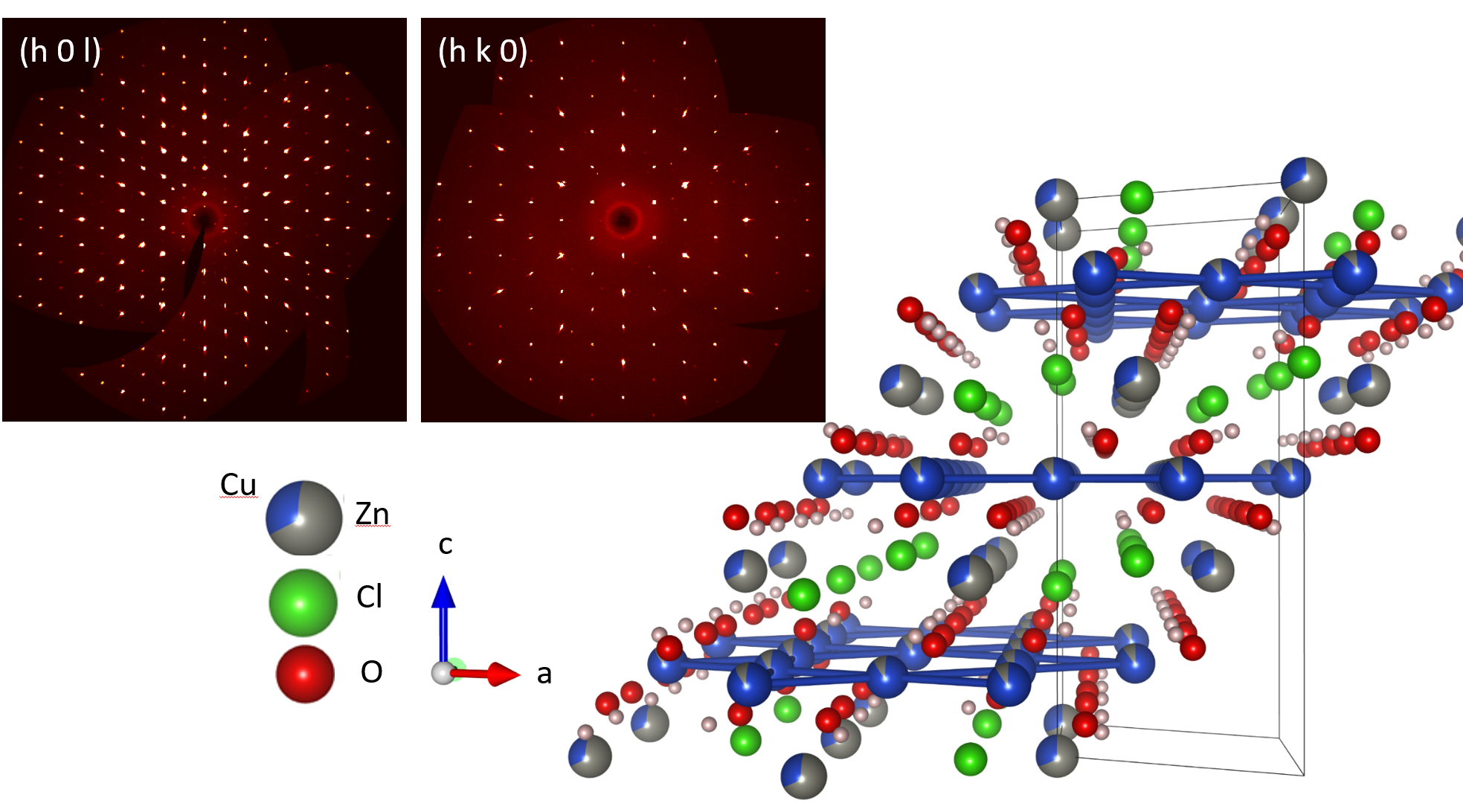}
\caption{Single crystal XRD zonal diffraction maps of the Zn$_{0.95(1)}$Cu$_{2.99(3)}$O$_{5.9(1)}$H$_{5.8(1)}$Cl$_2$ crystal and the resulting crystal structure viewed in a perspective mode from the single crystal data refinement ($R=1.22$) revealing a substitution of 10.8\% nonmagnetic Zn on the Cu sites. 
}
\label{Zn}
\end{figure}

Figure~\ref{Zn} shows zonal diffraction maps of this crystal and the refined crystal structure. A summary of the single crystal structure refinement is compiled in Table \ref{Zntab}.

\begin{table}
\caption{Crystal structure data of the Zn$_{0.95(1)}$Cu$_{2.99(3)}$O$_{5.9(1)}$H$_{5.8(1)}$Cl$_2$ crystal refined assuming the space group $R\overline3m$ (no. 166) with  $a$~=~$b$~=~6.8446(4)~\AA, $c$~=~14.0935(13)~\AA\ and $R$~=~0.012. In a separate table atom distances between the listed atom 1 and atom 2 are given,
}
\label{Zntab}
{\footnotesize{}}
\begin{tabular}{l|llllll}
\multicolumn{1}{l}{{\scriptsize{}atom}} & {\scriptsize{}x} & {\scriptsize{}y} & {\scriptsize{}z} & {\scriptsize{}$U_{\rm iso}$ (\AA$^2$)} & {\scriptsize{}occ.} & {\scriptsize{}Wyck.}\tabularnewline
\hline 
{\scriptsize{}Zn1} & {\scriptsize{}0} & {\scriptsize{}0} & {\scriptsize{}0} & {\scriptsize{}0.01153(8)} & {\scriptsize{}0.6753(9)} & {\scriptsize{}3a}\tabularnewline
{\scriptsize{}Cu1} & {\scriptsize{}0} & {\scriptsize{}0} & {\scriptsize{}0} & {\scriptsize{}0.01153(8)} & {\scriptsize{}0.3247(9)} & {\scriptsize{}3a}\tabularnewline
{\scriptsize{}Cu2} & {\scriptsize{}0.5} & {\scriptsize{}0} & {\scriptsize{}0.5} & {\scriptsize{}0.0112(3)} & {\scriptsize{}0.8918(3)} & {\scriptsize{}9d}\tabularnewline
{\scriptsize{}Zn2} & {\scriptsize{}0.5} & {\scriptsize{}0} & {\scriptsize{}0.5} & {\scriptsize{}0.017(3)} & {\scriptsize{}0.1082(3)} & {\scriptsize{}9d}\tabularnewline
{\scriptsize{}Cl} & {\scriptsize{}0} & {\scriptsize{}0} & {\scriptsize{}0.30510(3)} & {\scriptsize{}0.01618(10)} & {\scriptsize{}1} & {\scriptsize{}6c}\tabularnewline
{\scriptsize{}O} & {\scriptsize{}0.53930(7)} & {\scriptsize{}0.46070(7)} & {\scriptsize{}0.22816(5)} & {\scriptsize{}0.01392(14)} & {\scriptsize{}1} & {\scriptsize{}18h}\tabularnewline
{\scriptsize{}H} & {\scriptsize{}0.477(3)} & {\scriptsize{}0.523(3)} & {\scriptsize{}0.2461(18)} & {\scriptsize{}0.039(8)} & {\scriptsize{}1} & {\scriptsize{}18h}\tabularnewline
\end{tabular}{\footnotesize \par}

{\scriptsize{}}%
\begin{tabular}{cc|c}
{\scriptsize{}atom 1} & {\scriptsize{}atom 2} & {\scriptsize{}distance (\AA)}\tabularnewline
\hline 
{\scriptsize{}Zn1/Cu1} & {\scriptsize{}O} & {\scriptsize{}2.1160(8) }\tabularnewline
{\scriptsize{}Cu2/Zn2} & {\scriptsize{}O} & {\scriptsize{}1.9879(4)}\tabularnewline
{\scriptsize{}Cu2/Zn2} & {\scriptsize{}Cl} & {\scriptsize{}2.7768(3)}\tabularnewline
\end{tabular}{\scriptsize \par}

\end{table}

\begin{figure}[h]
\centering
\includegraphics[width=1.0\columnwidth]{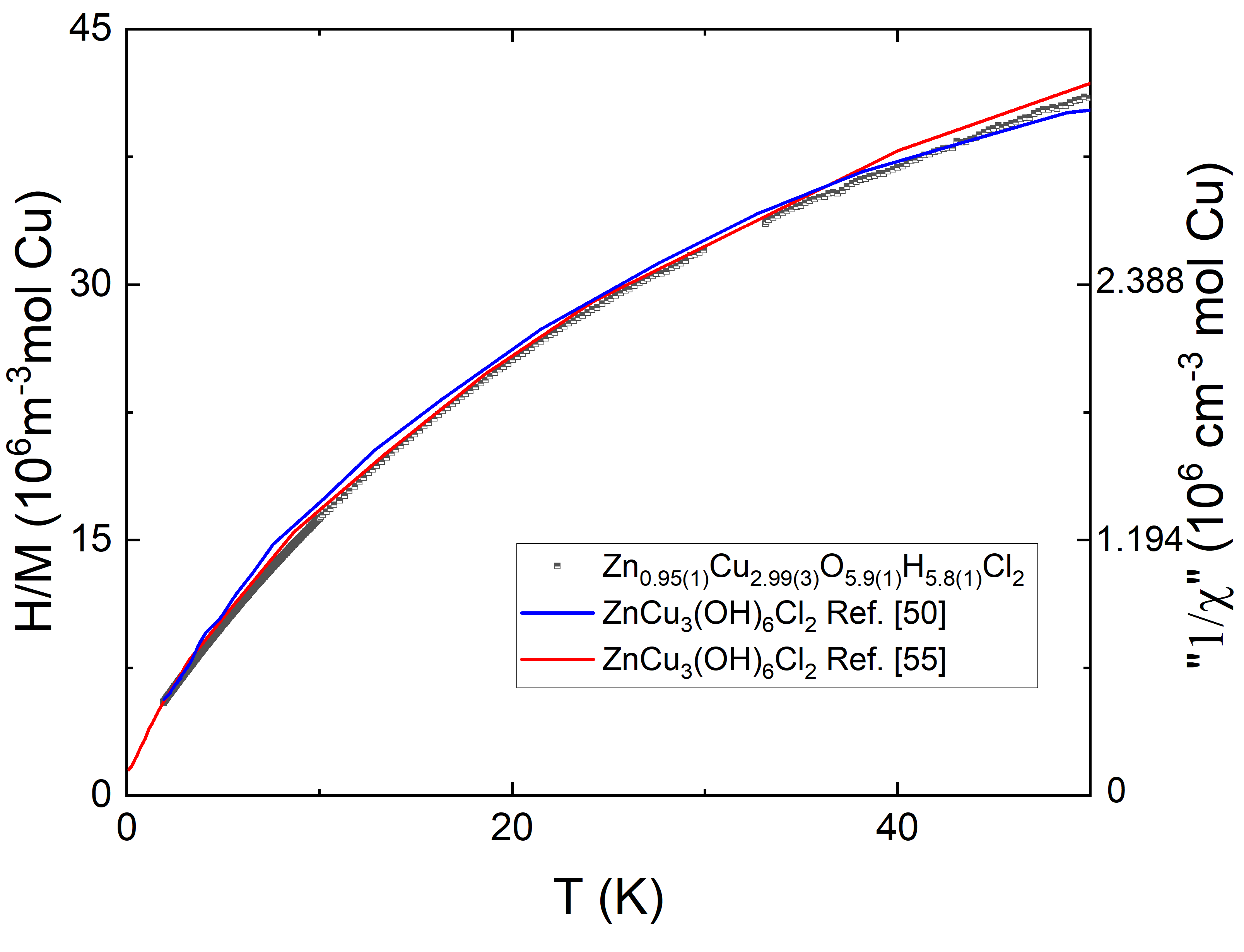}
\caption{Inverse $M/H$ or as a function of temperature of our x~=~1 herbertsmithite ZnCu$_3$(OH)$_6$Cl$_2$ crystal compared to results of two other groups \cite{Bert2007,Han2011} (reuse with permission).
Note: The data by Bert \textit{et al.} given in cgs units were converted to SI units and hence mulitplied by 4$\pi$ and the data by  Han \textit{et al.} by an additional correction factor of 1.18.
}
\label{Znsusc}
\end{figure}

The crystal structure of our crystal with composition Zn$_{0.95(1)}$Cu$_{2.99(3)}$O$_{5.9(1)}$H$_{5.8(1)}$Cl$_2$ was solved assuming the established space group $R\overline3m$ (no. 166) as given by Braithwaite \textit{et al.} and Shores \textit{et al.}\cite{Braithwaite2004,Shores2005}.
First we refined the structure placing Zn on the expected interlayer Wyckoff site 3a and Cu on the kagom\'e site 9d reaching a satisfying refinement with $R$~=~0.0129, $wR$~=~0.0315, and a Goodness-of-fit, GOF~=~1.173. 

To consider possible Cu:Zn site mixing  we allowed  the Zn atoms to occupy the kagom\'e Cu site (Wyckoff sites 9d)  and Cu atoms to substitute for the interlayer Zn atoms (Wyckoff site 3a). In view of the result of the chemical analysis we fixed the ratio of the substituted Zn:Cu atoms to 1~:~3. With this constraint we find 10.8\% of nonmagnetic Zn ions in the kagom\'e planes. Refinement of the crystal structure allowing for site mixing
improved the reliability factors to  $R$~=~0.0122, $wR$~=~0.0297, and the Goodness-of-fit to GOF~=~1.115. 
We note that the very similar xray scattering factors of Zn and Cu naturally make it difficult to locate the respective atom distribution with absolute certainty.
However, our result gains some support from the powder neutron diffraction investigation by de Vries \textit{et al.} who asserted very similar Zn:Cu mixing ratios \cite{Vries2008}. 
We also emphasize that our investigated Zn$_{0.95(1)}$Cu$_{2.99(3)}$O$_{5.9(1)}$H$_{5.8(1)}$Cl$_2$ crystal showed no twinning  which  significantly alleviated and simplified the structure refinement reflected by the very low reliability factors.

The Zn atoms occupying  sites within the kagom\'e layers dilute the magnetic system and induce random local distortion and strain since for Zn with a filled 3$d$ electron shell  the distortion of the O~-~Cl first anion coordination shell may be altered whereas the 32.4\% Cu atoms on the 3a interlayer site tend to Jahn-Teller distort the coordinating oxygen octahedra a possibility already raised  by Han \textit{et al.}.\cite{Han2016} As Norman \textit{et al.} analyzed, the crystal structure of herbertsmithite is prone to symmetry lowering\cite{Norman2020}. Local strain due to the Zn:Cu site mixing assists and stabilizes a higher symmetry, as it pins down the structure impeding distortions of the kagom\'e plane from magneto-phonon coupling, which happens here as the parent compound Cu$_2$(OH)$_3$Cl already above room temperature \cite{Malcherek2017} moves from the trigonal herbertsmithite-type structure $R\overline{3}m$ to the monoclinic clinoatacamite-type structure $P2_1/n$ \cite{Malcherek2017}. Even further symmetry lowering was found at low temperatures \cite{Zheng2013}. Via Zn substitution the high temperature $R\overline{3}$ is stabilized down to lowest temperatures in herbertsmithite.

The amount of free Cu spins can be estimated via the low temperature part in the magnetic susceptibility following Curie-Weiss law. We have measured several herbertsmithite crystals and found a remarkable reproducibility of the susceptibility at low temperatures. In Figure \ref{Znsusc} we show the low temperature part of the inverse susceptibility or $H/M$ in SI units of our crystal compared to literature data  taken from Refs. \onlinecite{Bert2007,Han2011}. The data by Han \textit{et al.} were multiplied by a factor  1.18, as there was an apparent scaling difference likely from small errors in the sample  mass determination. The good congruence of the magnetic susceptibilities from different sources and the good agreement of our XRD single crystal refinement results with the neutron power diffraction findings points to an equilibrium Cu-Zn site mixing level throughout  herbertsmithite.

\subsection{YCu$_3$(OH)$_{6}$Cl$_{3}$ and  Y$_3$Cu$_9$(OH)$_{19}$X$_{8}$ (X=Cl, Br)} 
\begin{figure}[h]
\centering
\includegraphics[width=1.0\columnwidth]{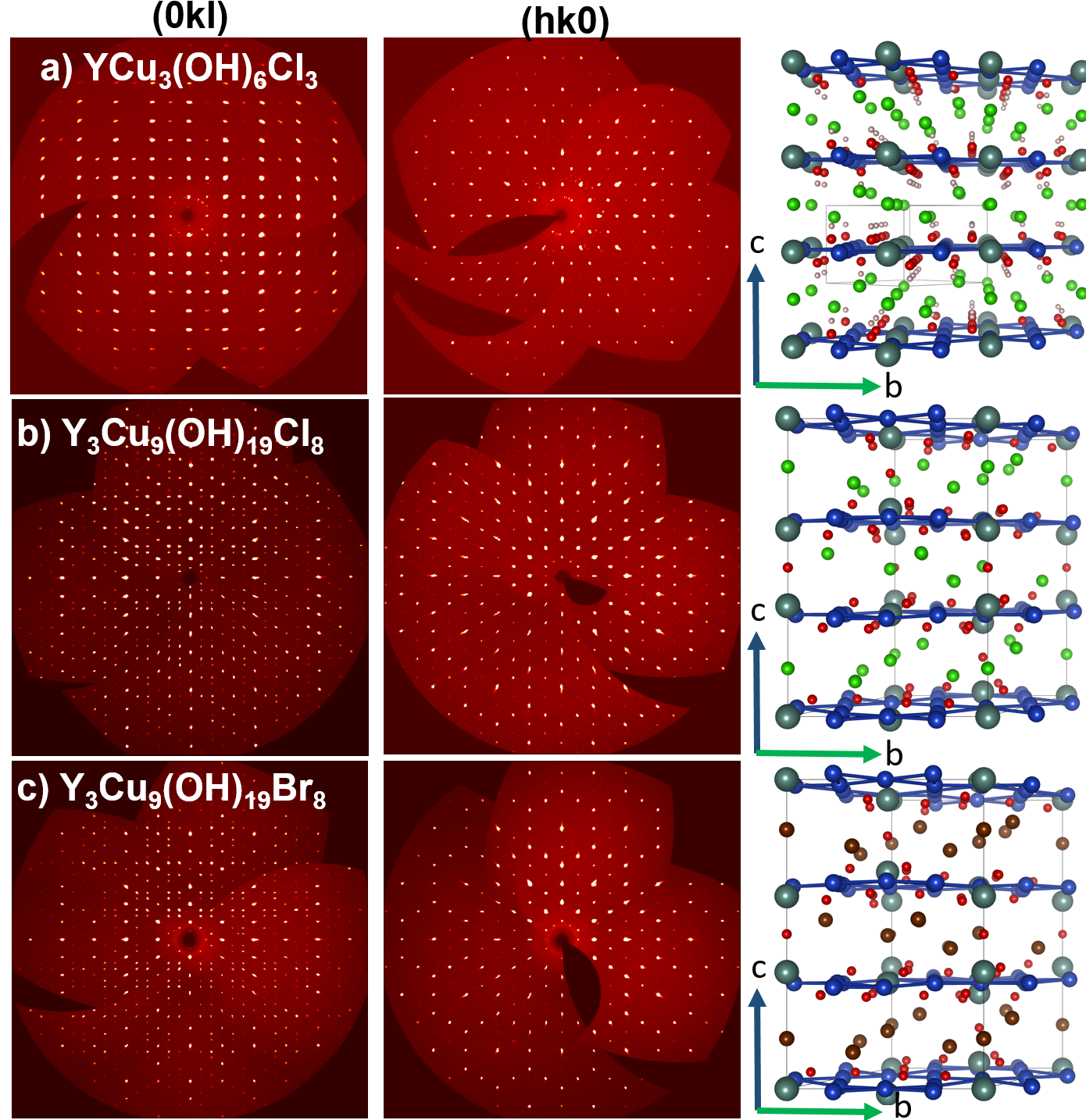}
\caption{Zonal  XRD maps of single crystals of a) YCu$_3$(OH)$_{6}$Cl$_{3}$ ($R$=0.0358, $GOF$=1.018), b) Y$_3$Cu$_9$(OH)$_{19}$Cl$_{8}$ ($R$=0.0339, GOF=1.028), and c) Y$_3$Cu$_9$(OH)$_{19}$Br$_{8}$ ($R$=0.038, GOF=1.04) as well as their crystal structure in a perspective view projected along $a^*$ as obtained from our single crystal structure refinements.
Cu atoms are displayed in blue,  Y atoms in metallic grey, and Cl and Br atoms in green and brown, respectively
}
\label{Y}
\end{figure}

The example of herbertsmithite raises the question, as to the role of disorder  in the YCu$_3$(OH)$_{6}$Cl$_{3}$ and  Y$_3$Cu$_9$(OH)$_{19}$X$_{8}$ (X=Cl, Br) kagom\'e systems.
We therefore investigated in detail their crystal structure by single crystal XRD.
To start with  we initially performed single crystal XRD on crystals of YCu$_3$(OH)$_{6}$Cl$_{3}$  and subsequently on Y$_3$Cu$_9$(OH)$_{19}$Cl$_{8}$. Our structure analysis proves the space group $P\overline{3}m1$ (no. 164) for YCu$_3$(OH)$_{6}$Cl$_{3}$, as reported before\cite{Sun2016}. In addition, weak superstructure Bragg reflections (see Fig.~\ref{Y}b) are seen for Y$_3$Cu$_9$(OH)$_{19}$Cl$_{8}$ which violate the selection rules for the higher symmetry space group  $P\overline3m1$, but can be readily explained by performing a refinement in space group $R\rm{\bar3}$ (no. 148) with a tripled unit cell. \cite{Puphal2017} Notably, $R\rm{\bar3}$ is a structural subgroup of $P\overline3m1$.

\begin{table}
\caption{Single crystal XRD results (atom fractional coordinates, atom displacement parameters $U_{\rm iso}$, occupancies)  obtained for a crystal with composition Y$_3$Cu$_9$(OH)$_{19}$Br$_{8}$. The refinement was carried out assuming the space group  $R\rm{\bar3}$ (no.~148) and the lattice parameters converged to  $a=b=11.5945(5)$\AA, $c=18.2302(11)$\AA. In a separate table the atom distances between the listed atom 1 and atom 2 are given,
}
\label{Ytab}
\begin{tabular}{l|lllll}
\multicolumn{1}{l}{{\scriptsize{}Site}} & {\scriptsize{}x} & {\scriptsize{}y} & {\scriptsize{}z} & {\scriptsize{}$U_{\rm iso}$ (\AA$^2$)} & {\scriptsize{}occ.}\tabularnewline
\hline 
{\scriptsize{}Br1} & {\scriptsize{}0} & {\scriptsize{}0} & {\scriptsize{}0.83774(2)} & {\scriptsize{}0.01823(9)} & {\scriptsize{}1}\tabularnewline
{\scriptsize{}Br2} & {\scriptsize{}0.33194(2)} & {\scriptsize{}0.00211(2)} & {\scriptsize{}0.78497(2)} & {\scriptsize{}0.01895(7)} & {\scriptsize{}1}\tabularnewline
{\scriptsize{}Y1} & {\scriptsize{}0} & {\scriptsize{}0} & {\scriptsize{}0.62289(2)} & {\scriptsize{}0.01134(8)} & {\scriptsize{}1}\tabularnewline
{\scriptsize{}Y2} & {\scriptsize{}0} & {\scriptsize{}0} & {\scriptsize{}0} & {\scriptsize{}0.01351(10)} & {\scriptsize{}1}\tabularnewline
{\scriptsize{}Cu1} & {\scriptsize{}0.17073(3)} & {\scriptsize{}-0.15827(3)} & {\scriptsize{}0.66160(2)} & {\scriptsize{}0.01271(7)} & {\scriptsize{}1}\tabularnewline
{\scriptsize{}Cu2} & {\scriptsize{}0.5} & {\scriptsize{}0} & {\scriptsize{}0} & {\scriptsize{}0.01271(8)} & {\scriptsize{}1}\tabularnewline
{\scriptsize{}O1} & {\scriptsize{}0} & {\scriptsize{}0} & {\scriptsize{}0.5} & {\scriptsize{}0.0404(15)} & {\scriptsize{}1}\tabularnewline
{\scriptsize{}O2} & {\scriptsize{}0.00241(19)} & {\scriptsize{}-0.17029(16)} & {\scriptsize{}0.69711(10)} & {\scriptsize{}0.0129(3)} & {\scriptsize{}1}\tabularnewline
{\scriptsize{}O3} & {\scriptsize{}0.3427(2)} & {\scriptsize{}-0.13493(18)} & {\scriptsize{}0.62560(10)} & {\scriptsize{}0.0125(3)} & {\scriptsize{}1}\tabularnewline
{\scriptsize{}O4} & {\scriptsize{}0.20379(17)} & {\scriptsize{}0.00614(19)} & {\scriptsize{}0.61068(9)} & {\scriptsize{}0.0124(3)} & {\scriptsize{}1}\tabularnewline
\end{tabular}

{\scriptsize{}}%
\begin{tabular}{ll|l}
{\scriptsize{}atom 1} & {\scriptsize{}atom 2} & {\scriptsize{}distance (\AA)}\tabularnewline
\hline 
{\scriptsize{}Cu1} & {\scriptsize{}O2} & {\scriptsize{}2.0047(18)}\tabularnewline
{\scriptsize{}Cu1} & {\scriptsize{}O4} & {\scriptsize{}1.9916(18)}\tabularnewline
{\scriptsize{}Cu1} & {\scriptsize{}Br2} & {\scriptsize{}2.8741(3)}\tabularnewline
{\scriptsize{}Cu2} & {\scriptsize{}O2} & {\scriptsize{}1.9938(18)}\tabularnewline
{\scriptsize{}Cu2} & {\scriptsize{}O3} & {\scriptsize{}1.981(2)}\tabularnewline
{\scriptsize{}Cu2} & {\scriptsize{}O4} & {\scriptsize{}1.9778(18)}\tabularnewline
{\scriptsize{}Cu2} & {\scriptsize{}Br2} & {\scriptsize{}2.8974(4)}\tabularnewline
\end{tabular}{\scriptsize \par}

\end{table}
Next, we performed single crystal XRD on a crystal with nominal composition YCu$_3$(OH)$_{6}$Br$_{2}$[Br$_x$(OH)$_{1-x}$].  Zonal diffraction maps and the crystal structure are depicted in Figure \ref{Y} c). Again clear superstructure Bragg reflections are visible, similar to the Cl homologue, indicating a symmetry reduction from $P\overline{3}m1$. A structure description in space group $R\rm{\bar{3}}$ (no. 148) proved successful with the refinement results summarized in Table \ref{Ytab}. 
The apparent very similar crystal structure refinement results of the  Cl and Br crystals suggests the composition of the Br homologue accordingly as  Y$_3$Cu$_9$(OH)$_{19}$Br$_{8}$.

Notably our synthesis of the Y$_3$Cu$_9$(OH)$_{19}$Br$_{8}$ crystals did not differ essentially from that of Ref. \onlinecite{Chen2020}. However, we found in our growth batches both chemically disordered (typically optically intransparent) and chemically ordered Y-Kapellasite-type crystals. This finding suggests that for the published crystal structure those Y atoms which moved out of the kagom\'e plane will locally distort the kagom\'e lattice and hence release frustration. We conclude that there  is simply a mixture of domains with distorted and undistorted local structures, averaging to the partially occupied scenario.

The level of  distortion of the kagom\'e lattice is clearly reflected by the lowest Cu-Cu-Cu angles. For Y$_3$Cu$_9$(OH)$_{19}$Cl$_{8}$ the Cu-Cu-Cu angle amounts to 176.12(3)$\degree$. Notably, this angle is further lowered for Y$_3$Cu$_9$(OH)$_{19}$Br$_{8}$ to 174.71(3)$\degree$. 

The  exchange interactions were calculated for the Cl variant and a rather low $J'$ was found for the shortest Cu-Cu bond, Cu1~-~Cu2 of 3.2582(6)	\AA~\cite{Hering2022}.  The corresponding Cu1~-~O~-~Cu2 bond angle amounts to 110.93(13)$\degree$, whereas the elongated Cu2~-~O~-~Cu1 and Cu1~-~O~-~Cu1 bonds enclose an angle of $\approx$118$\degree$. 
As can be expected for the larger anion Br$^-$, we find a subtly elongation of this short Cu1~-~Cu2 bond to 3.2640(4)\AA for the Br homologue, while the bonding angle with 110.63(9)$\degree$ remains nearly unchanged. 
Interestingly, the Cu2~-~O~-~Cu1 bond is strongly shortened and encloses an angle of 113.9(7)$\degree$, whereas the longest Cu2~-~O~-~Cu2 bond remains relatively large with a bonding angle of 116.4(7)$\degree$. As the antiferromagnetic superexchange is strongest close to 180$\degree$ this hints toward a movement in direction of the QSL phase in the phase diagram given by Hering \textit{et al.} \cite{Hering2022}.

\begin{figure}[h]
\centering
\includegraphics[width=1.0\columnwidth]{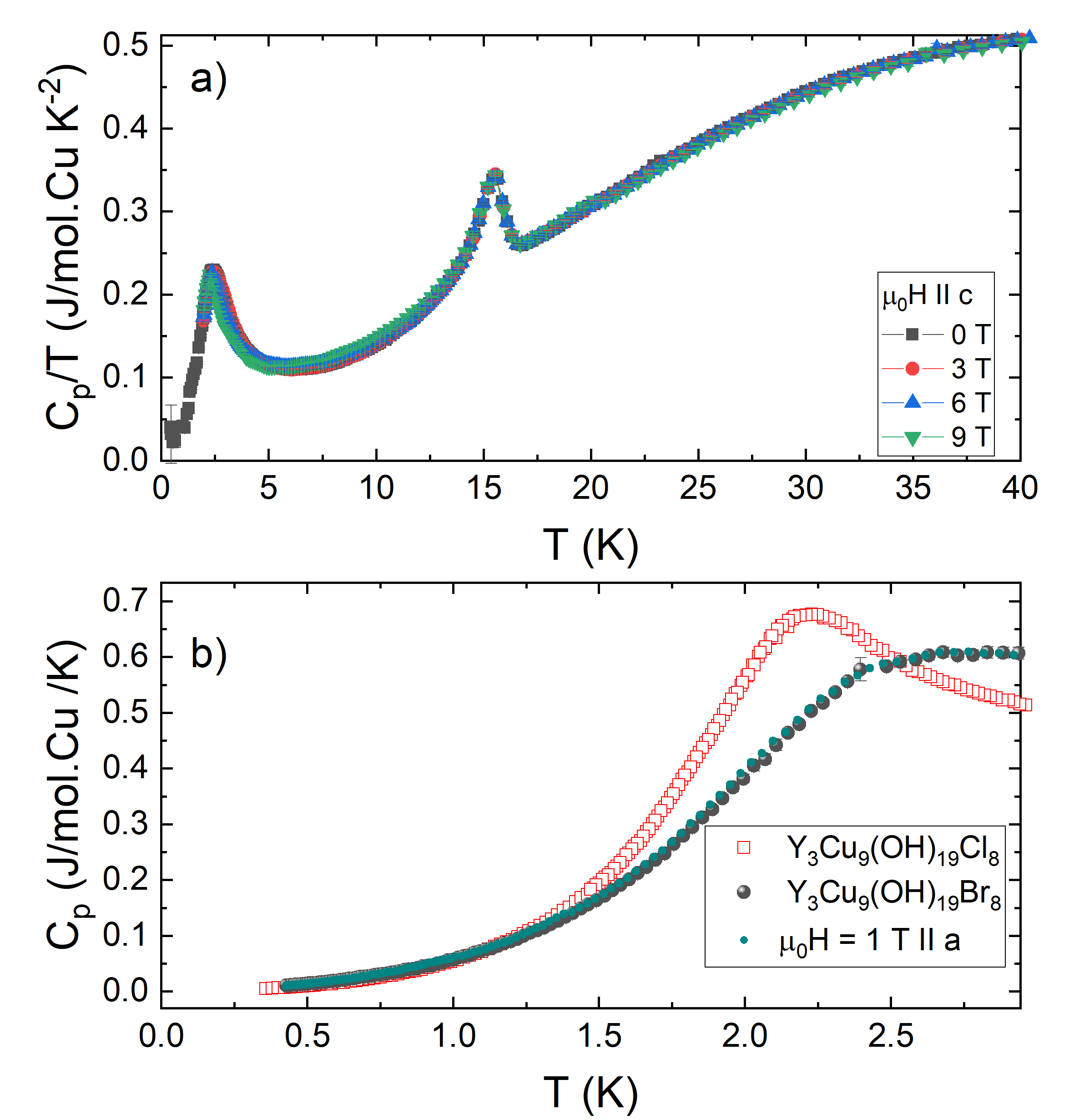}
\caption{a) Specific heat versus temperature of nine co-aligned Y$_3$Cu$_9$(OH)$_{19}$Br$_{8}$ single crystals with a total mass of 1.66 mg. b) Low temperature specific heat part of Y$_3$Cu$_9$(OH)$_{19}$Br$_{8}$ compared to our data of Y$_3$Cu$_9$(OH)$_{19}$Cl$_{8}$ \cite{Puphal2017}.}
\label{Cp}
\end{figure}
Naturally, to understand the physical properties of a system, it is best to investigate the chemically 'clean' phase-pure scenario. Hence, we characterized our Y$_3$Cu$_9$(OH)$_{19}$Br$_{8}$ crystals and measured their specific heat in the range temperature range from 0.4 to 40~K depicted in Figure \ref{Cp} a). Quite similar to literature and the Cl case we find a field-dependent maximum ataround 2.25~K indicating a magnetic transition also confirmed by spin wave measurements  and multi-magnon excitations \cite{Chatterjee2023,Biesner2022}. However, unlike literature, we observe a relatively sharp (in comparison to the magnetic transition in YCu$_3$(OH)$_{6}$Cl$_{3}$) field-independent peak at 15~K. For Y$_3$Cu$_9$(OH)$_{19}$Cl$_{8}$ there is only the 33~K anomaly in specific heat \cite{Chatterjee2023} (not shown here), while in thermal expansion and NMR a clear transition is also seen at the same temperature of 15~K \cite{Chatterjee2023}. In summary, this anomaly at 15~K likely originates from structural origin. The  $T^2$ slope of the specific heat in the narrow temperature range of 0~$<T<$~0.75~K below the 2.5~K peak   was attributed to a Dirac-like $T^2$ scenario.\onlinecite{Hong2022,Liu2022} However, such slope is equally well described by spin waves excitations and does not differ essentially from the behavior of the Cl homologue in this temperature range.

\begin{figure}[h]
\centering
\includegraphics[width=1.0\columnwidth]{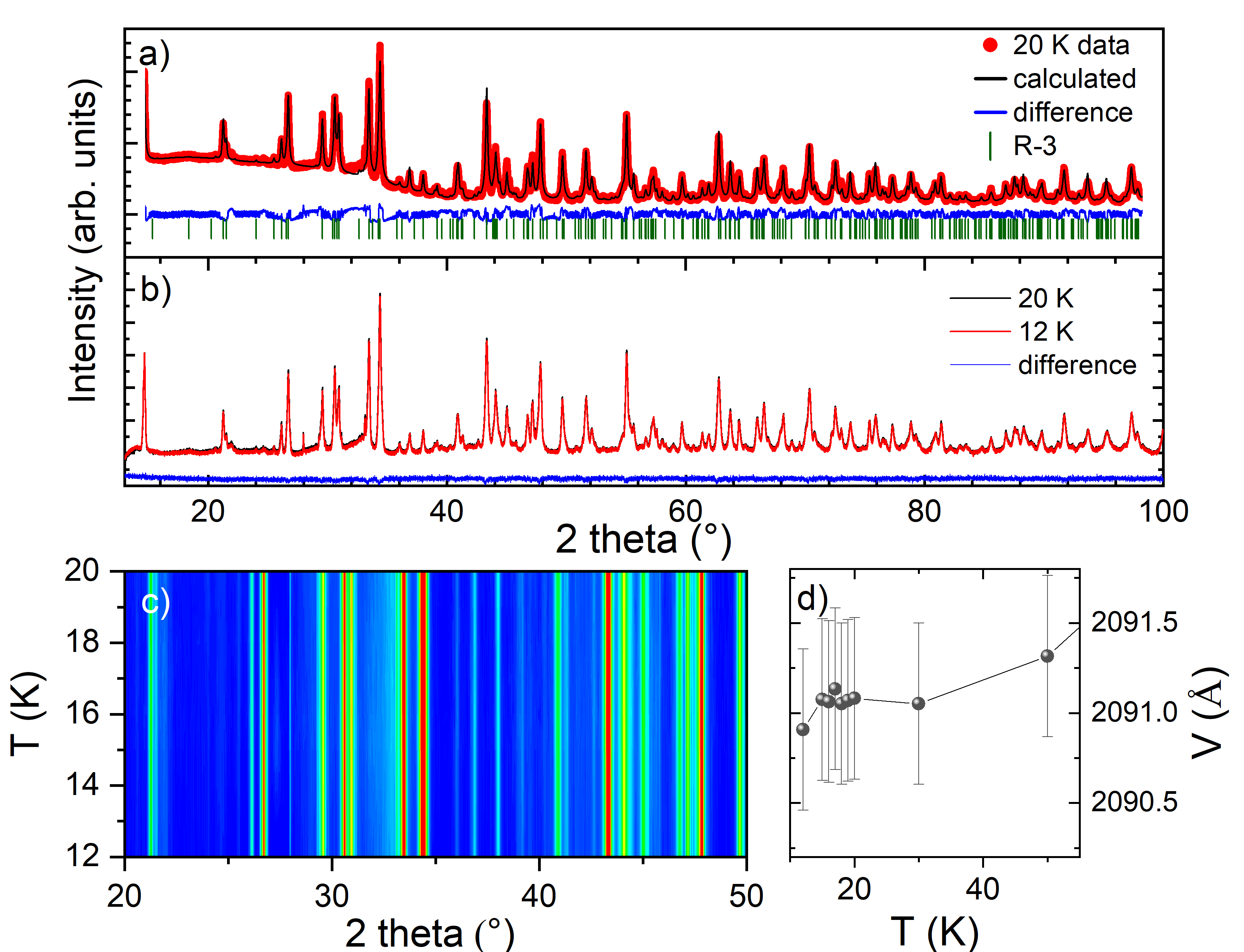}
\caption{(a) 12~K PXRD pattern (red) as well as the Rietveld profile refinement assuming the $R\overline{3}$ structure. (b) 20~K (black) and 12~K (red) PXRD pattern and the difference between them (blue). (c) Contour plots of a selected range of the PXRD data from 20-15~K revealing no major structural changes captured by PXRD. (d) Volume extracted from Rietfeld refinements versus temperature.
}
\label{lowTdiff}
\end{figure}

To investigate the low-temperature transition around 15~K we have carried out low-temperature powder XRD (PXRD) from 300 to 12 K, in a Bragg-Brentano setup, where the powder was first only spread on the sample holder with ethanol. At the structural transition due to a sudden volume change the sample jumped and we saw a zeroshift looking like a cell contraction. Notably, this is well in line with thermal expansion measurements on the Cl counterpart \cite{Chatterjee2023}. We then remeasured the sample mixed with ApiezonN grease. A selection of the PXRD data is shown in Figure \ref{lowTdiff} where no sudden symmetry reduction is observed as no peak splittings, nor sudden changes of the peak positions are visible. The transition certainly is of structural origin as besides our observation of sample jumping at exactly 18~K, a tiny anomaly in the crystal volume from Rietveld refinements and a deviation of the typical L-shape of the Volume change (see Figure \ref{lowTdiff} d), we find a peak splitting in Br-NMR \cite{Chatterjee2023thesis}. Hence, we have confirmed a similar subtle structural transition in the Br counterpart as in the Cl, which cannot be captured by simple PXRD, which likely stems from subtle hydrogen movements influenced by the magnetism. It is reasonable to assume that the system releases its magnetic frustration via a subtle structural distortion, notably the supercell and Cu distortion of two sites anyways is large enough to place us in the LRO scenario of the magnetic phase diagram \cite{Hering2022} and hence the system is likely finally resulting in LRO around 2~K very similar to its Cl counterpart \cite{Chatterjee2023}. In contrast the site disordered system shows no 15~K transition in specific heat \cite{zeng2022possible,Liu2022}.

\begin{figure}[tb]
\centering
\includegraphics[width=1.0\columnwidth]{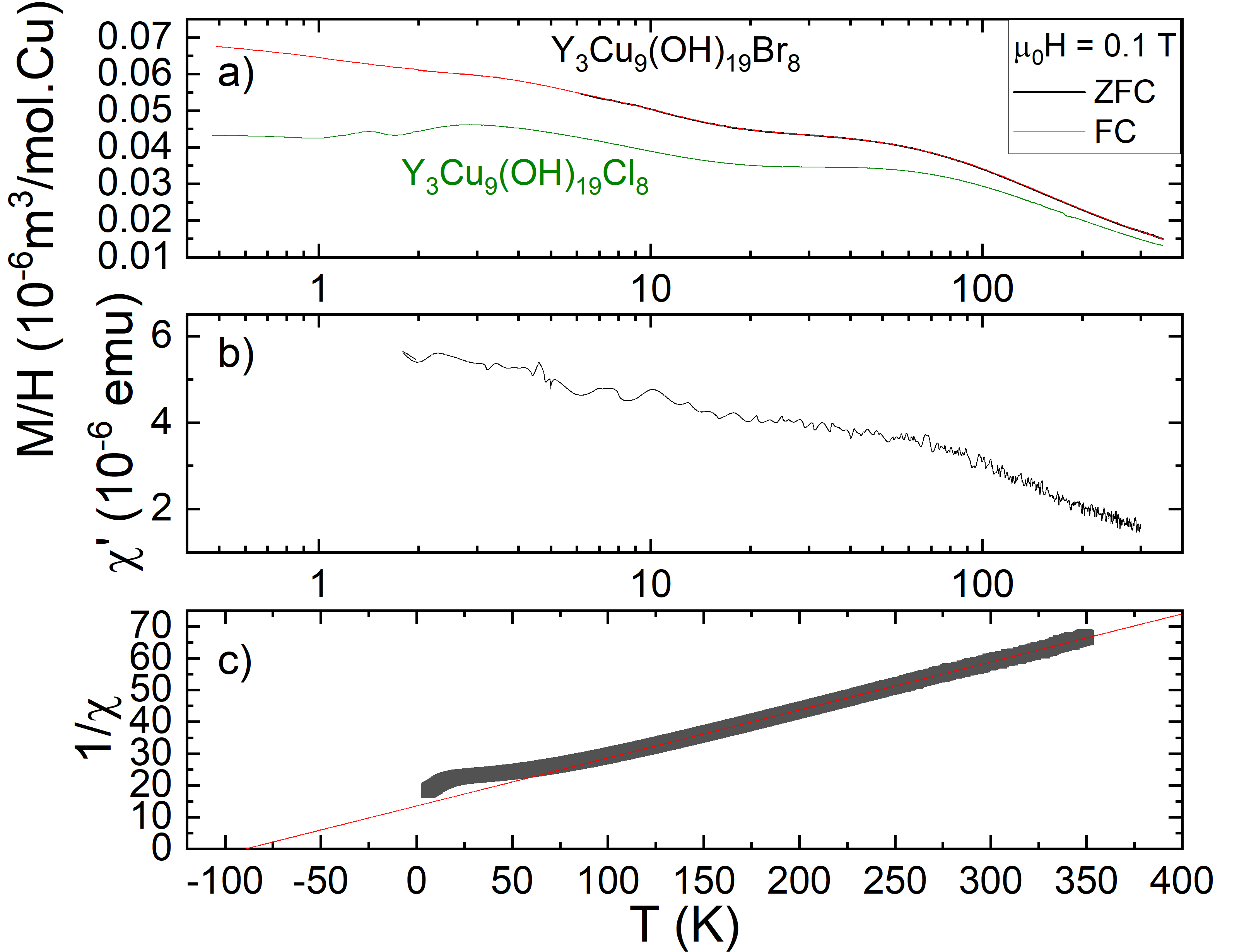}
\caption{a) Magnetization versus temperature on a semi-log plot of a collection of crystals of Y$_3$Cu$_9$(OH)$_{19}$Br$_{8}$ with a mass of 63~mg  (zero field cooled: black, field cooled: red)  compared to our data of Y$_3$Cu$_9$(OH)$_{19}$Cl$_{8}$ (green) \cite{Biesner2022}. b) AC susceptibility versus temperature on a semi-log plot. c) Inverse magnetization versus temperature fitted to a Curie-Weiss law.}
\label{susc}
\end{figure}

As shown in Figure \ref{susc} a) the magnetization versus temperature of Y$_3$Cu$_9$(OH)$_{19}$Br$_{8}$ looks  remarkably similar to that of the Cl homologue except for the slope below 2.5~K, where the Cl phase shows several maxima and the Br phase exhibits a continuous increase, mainly from impurity spins. AC susceptibility measurements (see Figure \ref{susc} b) only indicate a subtle increase on lowering the temperature, while a clear slope change is visible at around 15~K both in DC and AC, similar as for the Cl system.
A Curie-Weiss fit of the inverse susceptibility shown in Figure \ref{susc} c) yields a Curie-Weiss temperature of -90~K and a slope of 0.153 corresponding to an effective moment of 2~$\mu_B$, consistent with typical effective moments of Cu$^{2+}$ systems. All these findings are in good agreement with published results describing the disordered YCu$_3$(OH)$_{6}$Br$_{2}$[Br$_x$(OH)$_{1-x}$] phase\cite{Chen2020,zeng2022possible} and hint towards uniform physical properties rather insensitive to the disorder.

\begin{figure}[tb]
\centering
\includegraphics[width=1.0\columnwidth]{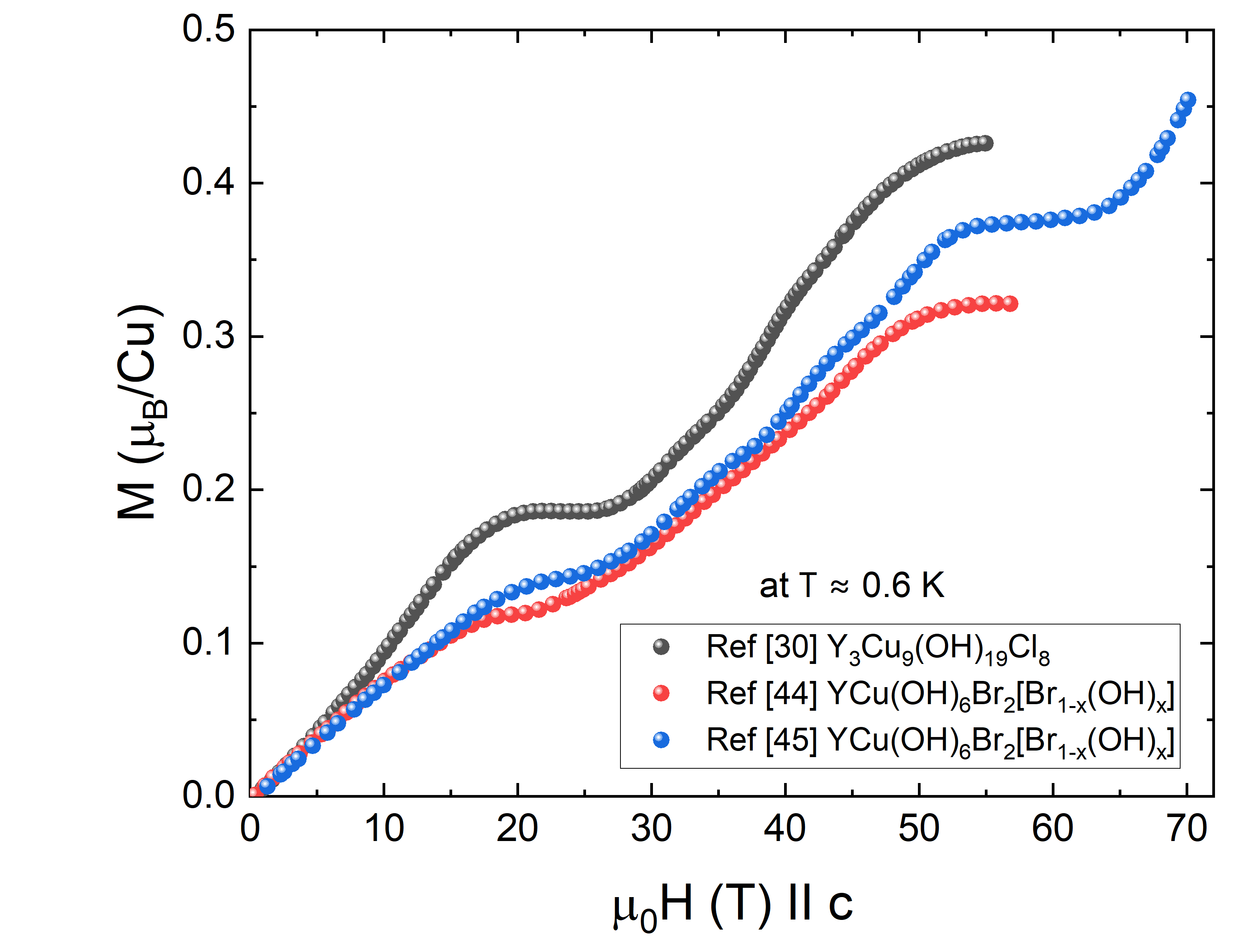}
\caption{Reproduction from literature of the magnetization versus magnetic field of our data on Y$_3$Cu$_9$(OH)$_{19}$Cl$_{8}$ \cite{Biesner2022} and data on YCu$_3$(OH)$_{6}$Br$_{2}$[Br$_x$(OH)$_{1-x}$] \cite{Suetsugu2024,Zheng2023} (reuse with permission) measured at $\sim$0.6 K for fields applied perpendicular to the kagom\'e plane.}
\label{MH}
\end{figure}

Finally, in  Figure \ref{MH} we compare the magnetizations as a function of the magnetic field collected at $\sim$0.6~K  of the two compounds Y$_3$Cu$_9$(OH)$_{19}$Cl$_{8}$ \onlinecite{Biesner2022} and YCu$_3$(OH)$_{6}$Br$_{2}$[Br$_x$(OH)$_{1-x}$] as reported in literature \cite{Biesner2022,Suetsugu2024,Zheng2023}.  Both compounds show magnetization plateaus at similar fields with the difference of the saturation value possibly originating from background signal, sample mass uncertainties and/or alignment issues as there is a sizable anisotropy. This already becomes apparent when comparing the two literature values of the same Br sample. Assuming that there are no such errors, Y$_3$Cu$_9$(OH)$_{19}$Cl$_{8}$ displays a 1/6 magnetization plateau and YCu$_3$(OH)$_{6}$Br$_{2}$[Br$_x$(OH)$_{1-x}$] a 1/9 plateau. However, given the structural similarities, similar physical properties can be expected. As we are in the same region of exchange interactions in the phase diagram proposed by Hering \textit{et al.} \onlinecite{Hering2022} we suggest that we observe similar magnetization plateaus in the two compounds.

In addition, a comparison of the INS  spectra of the two compounds Y$_3$Cu$_9$(OH)$_{19}$Cl$_{8}$ collected at 1.5~K (Ref. \onlinecite{Chatterjee2023}) and  YCu$_3$(OH)$_{6}$Br$_{2}$[Br$_x$(OH)$_{1-x}$] collected at 0.3~K (Ref. \onlinecite{Zeng2024}, is very suggestive that the same ground state is realized in both systems independent of disorder. The randomness induced by disorder simply increases the already quite broad distribution of $J'$ that exists in the Cl homologue as described by Chatterjee \textit{et al.}.\cite{Chatterjee2023} The scenario of randomness smearing out spin waves was well documented for Sr$_2$CuTe$_{1-x}$W$_x$O$_6$\cite{Fogh2022}. Via a continuous increase of randomness by substitution of Te by W it was demonstrated that the bond randomness, rather than the residual frustration, drives the physics of the system. The corresponding INS spectra simply smear out, while retaining their main features. However, we see a clear difference for the disorder free system as we find a structural transition at low temperatures, which is absent in the disordered crystals. Hence it is possible, that a spin liquid phase might be formed similar as in herbertsmithite.

\begin{figure}[tb]
\centering
\includegraphics[width=1.0\columnwidth]{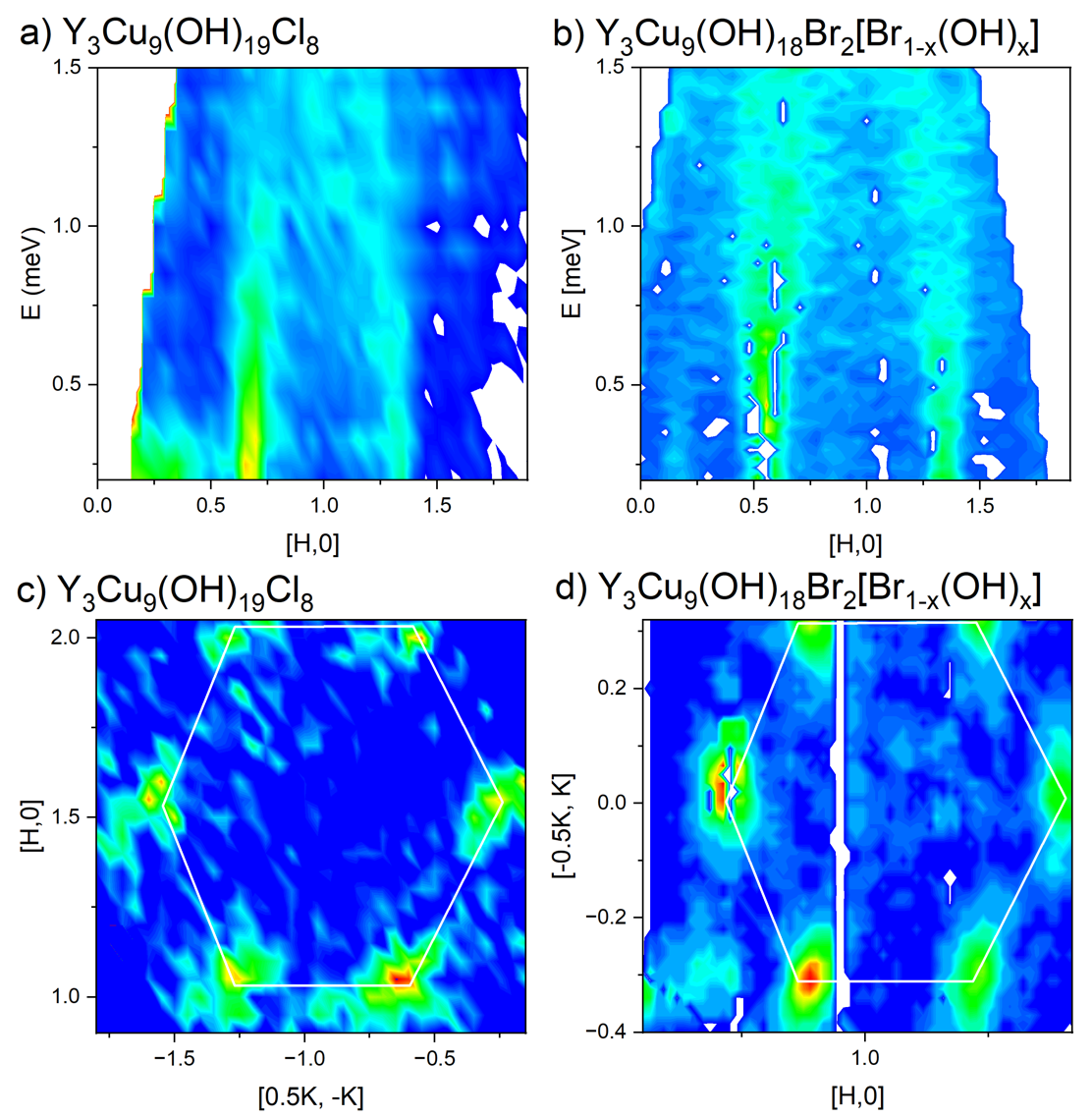}
\caption{Reproduction from literature of the inelastic neutron scattering data of a,c) Y$_3$Cu$_9$(OH)$_{19}$Cl$_{8}$ at 1.5~K \cite{Chatterjee2023} and b,d) YCu$_3$(OH)$_{6}$Br$_{2}$[Br$_x$(OH)$_{1-x}$] at 0.3~K \cite{Zeng2024} (note that the intensities are is in arbitrary units. In a,c) energy integrated plots are shown around 0.2 and 0.4~meV and in b,d) Intensity contour plots of the INS results are given as a function of $E$ and $Q$ along the [H,0] direction.}
\label{ins}
\end{figure}

\section{Summary}
In summary, we conclude that disorder clearly suppresses structural anomalies such as the 8~K transition of clinoatacamite in the Zn variant herbertsmithite, the 15~K transition in Y$_3$Cu$_9$(OH)$_{19}$Br$_{8}$ or the 33 K transition in Y$_3$Cu$_9$(OH)$_{19}$Cl$_{8}$. These structural transitions slightly distort the kagom\'e and release frustration leading to magnetic LRO. In the disordered systems signatures of LRO are less clear, as domains of distorted and undistorted kagom\'e layers co-exist preventing long spin waves to establish. As disorder appears to be a common feature in these systems it remains to be studied in more detail to what extent bond randomness is an essential ingredient  for the physics of  all existing QSL systems.
The chemo~-~structural disorder in the kagom\'e spin $S$~-~1/2 systems ZnCu$_3$(OH)$_6$Cl$_2$, found in a similar magnitude for independent samples and by different groups, leads to a substantial randomness with $\sim$11\% nonmagnetic spins in the kagom\'e plane. Similarly in  YCu$_3$(OH)$_{6}$Br$_{2}$[Br$_x$(OH)$_{1-x}$] due to the mixture of two structural variants with  $x$~=~0 and $x$~=~1/3.randomness is present for intermediate substitutions. The magnetic properties of our $x=1/3$  crystals resemble those of the Cl compound with a low-temperature structural transition emerging from strong magneto-elastic coupling as a release of frustration. The very close similarity of their physical properties of the two sister compounds suggests a (1/3,1/3) ground-state  possibly even for the disordered YCu$_3$(OH)$_{6}$Br$_{2}$[Br$_x$(OH)$_{1-x}$] phases.
\bibliography{kag}
\end{document}